\documentclass{nature}
\pdfoutput=1
\usepackage{color}
\usepackage{amsmath}
\usepackage{graphicx}
\usepackage[normalem]{ulem} 
\usepackage{lineno}
\newcommand{\MD}[1]{{{\color{black}#1}}} 
\newcommand{\modified}[1]{{{\color{black}#1}}} 
\newcommand{\fixed}[1]{#1} 

\title{Unstable fronts and stable ``critters'' formed by microrollers}

\author{Michelle Driscoll$^{1,\dagger,*}$, Blaise Delmotte$^{2,\dagger,*}$,  Mena Youssef$^3$, Stefano Sacanna$^3$, Aleksandar Donev$^{2}$ \& Paul Chaikin$^{1}$}

\begin{document}

\maketitle

\begin{affiliations}
 \item Department of Physics, New York University, New York, NY 10003, USA.
 \item Courant Institute of Mathematical Sciences,
New York University, New York, NY 10012, USA.
 \item Department of Chemistry, New York University, New York, NY 10003, USA.
 \item[$\dagger$] These authors contributed equally.
 \item[*] Corresponding authors.
\end{affiliations}

\begin{abstract}

Condensation of objects into stable clusters occurs naturally in equilibrium\cite{Anderson2002} and driven systems\cite{Aranson2006, Marchetti2013, Palacci2013, Bialke2015}.  It is commonly held that potential interactions\cite{Lu2013}, depletion forces\cite{Schwarz2012}, or sensing\cite{Vicsek2012} are the only mechanisms which can create long-lived compact structures.  Here we show that persistent motile structures can form spontaneously from hydrodynamic interactions \emph{alone} with no sensing or potential interactions.  We study this structure formation in a system of colloidal rollers suspended and translating above a floor, using both experiments and large-scale 3D simulations. In this system, clusters originate from a previously unreported fingering instability, where fingers pinch off from an unstable front to form autonomous ``critters'', whose size is selected by the height of the particles above the floor. These critters are a stable state of the system, move much faster than individual particles, and quickly respond to a changing drive. With speed and direction set by a rotating magnetic field, these active structures \MD{offer interesting possibilities for guided transport, flow generation, and mixing at the microscale.}

\end{abstract}

\newpage

We have identified a new instability in one of the most basic systems of low-Reynolds number (steady Stokes or overdamped) flow, a collection of spheres rotating near a wall.  This system has been well-studied analytically and numerically\cite{Blake1974,Lee1980}, since it is considered a base model for understanding many microbial and colloidal flows.  The instability visually resembles wet paint dripping down a wall or individual droplets sliding down a windshield\cite{Huppert1982}, examples of Rayleigh-Taylor instabilities\cite{Chandrasekhar1961}. However, in those and other clustering phenomena what holds things together is surface tension or other forces deriving from an interaction potential.  Here we use a model system to explore whether hydrodynamic interactions alone, without particle collisions, attractions or sense/response redirection can lead to stable finite clusters.  

The experimental system consists of polymer colloids with radius  $a=0.66$ $\mu$m which have a small permanent magnetic moment ($\lvert \vec{m}\rvert \sim 5\cdot10^{-16}$ Am$^2$) from an embedded hematite cube\cite{Sacanna2012}, see schematic in Fig.\ 1a. Inter-particle magnetic interactions are small compared to thermal energy ($< 0.1$ $k_BT$).  A rotating magnetic field ($\vec{B} = B_0\left[\cos(\omega t)\hat{x}+\sin(\omega t)\hat{z}\right]$) with magnitude $B_0$ and frequency $f=\omega/2 \pi$ is applied, causing all the particles to rotate about the $\hat{y}$-axis at the same rate $\omega$. \MD{  The particles rotate synchronously with the field for $\omega < \omega_c$, where $\omega_c$ is the critical frequency above which the applied magnetic torque is not enough to balance the viscous torque on the particle, see SI Section I for details of the rotation mechanism.  In all of our experiments, $\omega < \omega_c$.} In contrast with recent experiments on Quincke rollers\cite{Bricard2013}, the rotation direction is prescribed and does not arise from the system dynamics. Hydrodynamics is the dominant inter-particle interaction in this system, which is distinctly different from many other systems of rotating magnetic particles, where dynamics is found to be a strong function of inter-particle magnetic interactions\cite{Sing2010, Martinez2015,Grzybowski2000, Snezhko2016}. \MD{Many ferromagnetic particles with a small remnant moment could produce the same behaviour.} Gravity plays an unusual role in our instability. Rather than a driving force partially compensated by surface tension or viscosity, as  in Rayleigh-Taylor instabilities, here gravity is completely compensated by thermal motion and sets the average particle height, $h=a+k_BT/mg$, where $mg$ is the particle buoyant weight. The particles are contained in glass capillary chambers, with a depth $ \geq100$ $\mu$m. The particles have a density of 2,000 kg/m$^3$, and are thus suspended at $h \approx  1.0$ $\mu$m above the capillary floor, interacting essentially with only one wall.

 We model the system as particles near an infinite wall driven at a constant rotational frequency.  \modified{The many body simulations are done using an accelerated stresslet-free variant of Stokesian dynamics\cite{Swan2007} that represents each particle as a single regularized stokeslet and rotlet singularity. The hydrodynamic interactions between the particles, and the particles and the wall, are explicitly modeled (see Methods). This is a minimally-resolved and thus low-accuracy method. The accuracy of the numerical calculations can be improved by using a ``multi-blob'' approach where particles such as our rotating spheres are represented as a rigid cluster of regularized stokeslet singularities or ``blobs''\cite{Usabiaga2016}. The accuracy and resolution are set by the number of blobs per particle. Improved resolution comes at the computational expense of a reduction in the number of particles and the time span that can be simulated. In SI Figure 1b we compare a well-resolved sphere model containing 2562 blobs, which is used in Figure 1c and 1d, with the minimally-resolved model used elsewhere, which produces qualitatively similar results.}
 The simulations do not include the effects of thermal motion \MD{nor magnetic forces as they are small compared to viscous forces (see SI Section II).} The particle-wall separation is set by creating a repulsive wall potential.  In addition to the hydrodynamic interactions between particles, steric repulsion between particles is used for some simulations, e.g., to model the experiment, and dropped for others, e.g., to test the role of pure hydrodynamics.

Hydrodynamic coupling plays a crucial role in the dynamics of this system. A sphere rotating near a wall about the $\hat{y}$-axis will move in the $\hat{x}$-direction (see Fig.\ 1c).  
This motion is a result of the unequal drag force on the top and the bottom of the particle. Its translation speed $v_0$ is set by the scaled distance to the wall and the rotation rate\cite{Lee1980}: $v_0= \omega a f(h/a)$, and vanishes as the height increases, i.e. $f(h/a) \rightarrow 0$ as $h \rightarrow \infty$.   However, as shown in Fig.\ 1c, $v_0$ is orders of magnitude slower than the fluid velocity at the particle's surface $\omega a$. The velocity field around a particle decays slowly (as $1/r^2$) in the $xy$-plane, where $r$ is the distance to the particle centre.  Thus, even at moderate area fractions, $\phi$, particle motion is mainly a result of being advected in the flow of neighbouring particles.  This collective effect can be described by a density-dependent mean velocity, $\bar{v} \equiv (v-v_0)/v_0  = \alpha\phi$, where $\alpha$ is found to be \MD{$50 \pm 2$} in our experimental system, see Fig.\ 1b.  Putting $v \approx \alpha \phi v_0$  into the continuity equation,
\begin{equation}
    \frac{\partial \phi}{\partial t} + \frac{\partial (v\phi)}{\partial x}  =
    \frac{\partial \phi}{\partial t} + \alpha v_0 \frac{\partial \left( \phi ^2 \right) }{\partial x} = 0
\end{equation}
results in the inviscid Burger's equation, which is well-known to lead to the development of a shock front\cite{Burgers1974}.  Although the development of a shock is a simple consequence of having a density-dependent velocity, to our knowledge, this feature has only been observed in low-density flowing emulsion systems\cite{Beatus2009,Desreumaux2013}, and not other colloidal roller systems\cite{Bricard2013}.  As in Burger's solution, we need a leading edge density gradient to observe the shock.  

The propagating shock front quickly becomes unstable in the direction transverse to propagation, leading to the appearance of density fluctuations which continue to grow as fingers (see Fig.\ 2a and SI Movies 1 and 2).  This fingering instability does not occur in a planar Burger's shock\cite{Goodman1999}.  Both the experiments and simulations show a qualitatively similar evolution of the shock, the shock instability, and the fingering, at the same relative times. Despite its deceptively similar appearance, this fingering instability is distinct from other previously reported viscous\cite{Huppert1982}, granular\cite{Pouliquen1997, Aranson2006} and colloidal instabilities\cite{Pan2001, Lin2014, Wysocki2009}.  In a Rayleigh-Taylor-like instability, the fingering dynamics is controlled by a balance of viscous damping and a body-force driving term (like gravity).  In contrast, the instability wavelength in this system is independent of both viscosity and driving torque (i.e. rotation rate), see Fig.\ 2b, c. We define  the instability wavelength, $\lambda_{\max}$, as the wavelength associated with the fastest growing normal mode, as is typical for a linear instability, see \MD{SI Section III} for details.  The  control parameter for this instability, in both the experiments and the simulations, is $h$, as illustrated in Fig.\ 2d, e.  As shown in that figure, when the particle-wall distance increases, the instability wavelength increases linearly.  The numerical results are obtained by confining the particles to a plane parallel to the wall at a given height $h$, see SI Movie 3. As discussed later in the text, the same dynamics are observed in this configuration as in the fully 3D simulations. In the experiments, the height is adjusted by changing the solvent density and hence the colloids' buoyant weight $mg$.  Although we change the particle-wall separation in quite different ways in the experiments and simulations, in both cases $h$ is the key control parameter for $\lambda_{\max}$.

Due to the increased density in the shock region, the  fingertips are much denser and, due to collective effects, \MD{i.e.\ $\bar{v} = \alpha\phi$}, move much  faster than the rest of the suspension. In the simulations, if the particles are maintained high enough away from the wall, the fingertips break off to form self-sustained, compact clusters made of hundreds of particles, which we term ``critters'' (see Fig.\ 3a and SI Movie 4). These critters rotate around their centre of mass and translate with a speed 15,000 times faster than a single roller would at the same centre of mass height (Fig.\ 3b inset). Critters form a natural stable state of the system: they move at a constant speed, do not lose particles, and are not observed to dissolve (see Fig. 3b and SI Movie 4).  We further explore their stability by changing the direction (but not the magnitude) of $\omega$ periodically in time. As shown in Fig.\ 3c and SI Movie 5, the critters follow the prescribed circular trajectory.  Somewhat similar structures to the critters were obtained experimentally  when $h$ was increased (Fig.\ 2e, and SI Movie 6), although in the experiment critters continually lost some particles as they moved. In the simulations, the compact critters are extremely robust, suggesting they may be an attractor in the system dynamics -- similar critters appear regardless of the initial conditions (see SI Movie 7).

The velocity field in Fig.\ 1d suggests that the transverse instability of the shock originates from the lateral hydrodynamic attraction and repulsion in the $xy$-plane; this lateral flow shows the same qualitative features for a rigid sphere and for a point rotlet above a no-slip boundary\cite{Blake1974}.  To test the assumption that this is a planar instability, we simplify the system in our  simulations by restricting the rollers to a fixed plane above the wall.  This simplified system reproduces the instability: the shock forms, the transverse instability develops, and autonomous clusters with selected size detach and translate much faster than individual particles, as shown in Fig.\ 4a. We further remove {\em all} non-hydrodynamic effects and simplify the system to its bare minimum by simulating instead a collection of point rotlet singularities without any steric repulsion. Fig.\ 4c and SI Movie 8 show that \emph{both} the fingering instability and clustering are reproduced with only this one ingredient: hydrodynamic interactions in the vicinity of a no-slip boundary. 

A closer look at the flow field around a cluster in the frame moving with its centre of mass shows a well defined recirculation zone whose size matches the cluster size, as shown in Fig.\ 4b.  The closed streamlines in this flow field are responsible for the self-sustained and compact clusters. As seen in studies of sedimenting particle clouds\cite{Metzger2007}, the chaotic nature of the flow inside a cluster can lead to the loss of particles. However, in our 3D simulations, additional circulation in the $xz$-plane prevents particle loss; the critters are stabilised by the closed streamlines.
\MD{Critters smaller than the size of the recirculation zone, which is proportional to the height above the wall, can form and remain stable, while larger ones break up by shedding excess particles.}

Clustering is seen in many other low Reynolds number systems, from sedimentation to active colloidal particles.  What is \MD{notable} about the critters that emerge from this instability is that they are a persistent state which can be produced from hydrodynamic interactions.  Other kinds of hydrodynamic clusters, such as those seen in sedimentation\cite{Lowen2010,Guazzelli2011} are always transient and not long-lived structures.  Almost all active matter systems display some kind of clustering instability\cite{Marchetti2013,Palacci2013}, but it is usually a consequence of  particle-particle interactions, either directly through an attractive potential, sensing, or via self-trapping, which is a consequence of a repulsive particle potential.  Here we have demonstrated that the same instability observed in the experiments is preserved in the simulations, even when all interactions except hydrodynamics are removed. 

 In this study, we isolated the role of hydrodynamics. \MD{We note that this instability is generic and should be found in any system of particles rotating parallel to a floor, provided that hydrodynamics is the dominant particle-particle interaction. The addition of particle-particle potentials can strongly modify the instability structure and dynamics.} For example, varying the Mason number (relative strength of hydrodynamic and magnetic interactions) completely changes the dynamics of the system, see \MD{SI Section II and SI Movie 9}.  \MD{The structures and flow patterns formed in our model system suggest a number of possible applications. Collections of rollers create strong advective flows, and their motion and direction can be externally powered and controlled. As shown in SI Section IV, we have found, both experimentally and in simulations, a number of promising ways to transport passive particles by microrollers in either homogeneous suspensions, fingers, or critters.}

\newpage

\begin{methods}

\subsection{Experiments.}

The colloidal particles are TPM (3-methacryloxypropyl trimethoxysilane) spheres ($a$ = 0.66 $\mu$m) with hematite cubes embedded into them, see Sacanna \emph{et al.}\cite{Sacanna2012} for details of the synthesis.  Hematite is a canted anti-ferromagnet, thus the particles possess a small permanent moment, $\lvert \vec{m}\rvert \sim 5\cdot 10^{-16}$ Am$^2$, which can be oriented in an applied magnetic field.  The particles are dispersed in either water or aqueous glycerol solutions (dynamic viscosity $\eta = 1$ mPa$\cdot$s,  $\eta = 4$ mPa$\cdot$s, or $\eta$ = 10 mPa$\cdot$s).  To increase buoyancy, additional samples are created with particles dispersed in a 410 mM sodium polytungstate solution, with a small amount of TMAH added as an additional stabiliser (1.2\% v/v).

In all cases, the particles are contained in glass capillary tubes of depth 100 $\mu$m or greater (VitroCom VitroTubes$^{TM}$), which are sealed with UV epoxy (Norland NOA63).  To create the initial density gradient in particle concentration, the chambers are tilted so that particles gather to one side, then are laid flat to ensure a monolayer is formed as the initial condition.  Distance of this initial gradient to the vertical chamber wall does not affect the instability wavelength.

The rotating magnetic field is created using custom triaxial coils.   A bipolar current supply (KEPCO BOP 50-2M) is used to apply the current to the coils and create a rotating magnetic field.  The waveforms for the rotating field are generated using a DAQ (MCC USB-3101FS) controlled via Matlab$^{TM}$.  The field generated by the coils is measured with a Hall sensor (Ametes MFS-3A).  For all experiments described in this work, the magnitude of the field is 2.94 mT, and the frequency is varied from 0.2 -- 25 Hz.

All observations are made using a Nikon Ti-U inverted microscope. \MD{Roller velocity is calculated in two ways.  At low area fractions ($\phi < 0.1$), the velocity is computed using the instantaneous velocity calculated from particle-tracking\cite{trackpy}.  The velocity was computed for small segments of the individual particle tracks, and the results were then binned to calculate the mean roller velocity.  At high area fractions, individual particle velocities cannot be measured, and a custom python code was used to process the images and perform Particle Image Velocimetry (PIV) analysis.  The roller velocity was then taken to be the mean suspension velocity computed from the PIV analysis. Using a range of area fractions where both particle tracking and PIV could be used,  $\phi = 0.10-0.20$, we validated that the mean suspension velocity matched the individual particle velocity, i.e.\ when hydrodynamic collective effects are predominant, the mean suspension velocity is equivalent to the individual particle velocity.}

\subsection{Simulations.}

The flow fields around one roller shown in Fig.\ 1 are obtained by using the rigid multi-blob method developed by Usabiaga \emph{et al.}\cite{Usabiaga2016}. The surface of the roller is discretised with 2562 blobs which are rigidly connected with constraint forces. 
Pairwise hydrodynamic interactions between blobs are modelled with the Rotne-Prager-Blake tensor with wall corrections\cite{Swan2007}. 

The multi-particle simulations are performed using the Stokesian Dynamics method developed by Swan and Brady\cite{Swan2007}, omitting stresslets and thermal fluctuations (Brownian motion). In brief, the hydrodynamic response is computed by replacing each sphere with a regularised singularity (stokeslet and rotlet) and accounting for the \fixed{hydrodynamic interaction with the} wall  in an approximate but self-consistent way \fixed{by applying Rotne-Prager corrections to} the Blake image construction\cite{Blake1974}. \MD{This modelling only includes leading order corrections for the finite size of the particles to limit the computational cost required to simulate large numbers ($O(10^4)$) of particles. Even though this low resolution model overestimates the particle mobility, it remains physically consistent and its \modified{accuracy can be controlled and quantified.  A more resolved multi-blob model,\cite{Usabiaga2016} would increase the hydrodynamic accuracy but incur a higher computational cost. We compare both the minimally-resolved model\cite{Swan2007} and the well-resolved multi-blob model\cite{Usabiaga2016}} with experiments in SI Section I.}

As in the experiments, the rotation rate of the particles is prescribed. This is ensured by applying the appropriate torques $\mathbf{T}$ obtained by solving the following resistance problem\cite{Swan2007}
\begin{eqnarray}
  \mathbf{M}^{rr}\mathbf{T} = \boldsymbol{\Omega} - \mathbf{M}^{rt}\mathbf{F}
\end{eqnarray}
 where $\mathbf{M}^{rr}$ is the mobility matrix coupling the prescribed particle rotations $\boldsymbol{\Omega}$ to the unknown particle torques $\mathbf{T}$. $\mathbf{M}^{rt}$ is the mobility matrix coupling $\boldsymbol{\Omega}$ to the known external forces $\mathbf{F}$ acting on the particles, which are a combination of particle-particle and particle-wall repulsive forces and gravity.  Once the torques are obtained, the translational velocities $\mathbf{V}$ are found with the mobility relation 
 \begin{eqnarray}
\mathbf{V} =  \mathbf{M}^{tr}\mathbf{T} + \mathbf{M}^{tt}\mathbf{F}
\end{eqnarray}
where \fixed{$\mathbf{M}^{tr}=(\mathbf{M}^{rt})^T$} couples $\mathbf{V}$ to $\mathbf{T}$ and $\mathbf{M}^{tt}$ couples $\mathbf{V}$ to $\mathbf{F}$.  When the particles are restricted to a plane at fixed height $z = h$, forces and motion in the $\hat{z}$-direction are discarded.  Particle trajectories are integrated with the two-step Adams-Bashforth-Moulton predictor-corrector scheme.
The time step $\Delta t$ is chosen so that a particle does not travel more than 5$\%$ of its size per time step: $v\Delta t< 0.05a$. Typically, $\Delta t = 0.005$ s in most of the simulations.
Mobility-vector products and steric interactions are computed with PyCUDA on an Nvidia K40 GPU. The typical simulation time is 7 hours for 20,000 time iterations with $O(10^4)$ particles.

When included, steric interactions between the particles are modelled with a pairwise soft-core repulsive potential $U_{\text{part}}$ \fixed{of Yukawa type,}
$$U_{\text{part}}(r) = S_p\frac{a}{r}\exp\left(-\frac{r}{D_p}\right),$$
where $r$ is the centre-to-centre distance between particles, $S_p$ is the strength of the potential ($S_p = 2.43\cdot10^8 mga$) and \fixed{the interaction range is} $D_p=0.1a$.
Since the simulations do not include Brownian motion, in order to balance gravity forces and set the equilibrium height of the particles we use a repulsive potential from the wall 
$$U_{\text{wall}}(z)= S_w \frac{a}{z-a}\,\exp\left(-\frac{z-a}{D_w}\right), $$
where $z$ is height of the particle centre. The strength $S_w$ and the \fixed{range} $D_w$ are changed between the simulations to investigate the effect of the particle height on the instability: $S_w = 0.05$ -- $25.1mga$  and $D_w = 0.1a$ -- $7a$.
The total force on the particles $\mathbf{F}$ is given by the \fixed{gradient of the} combination of the repulsive potentials, $U_{\text{part}}$ and $U_{\text{wall}}$, and the gravitational potential $mgz$, where $m = 1.27\times 10^{-15}$ kg is the excess mass of a roller.

\end{methods}

\newpage

\begin{addendum}
 \item This work was supported primarily by the Gordon and Betty Moore Foundation through Grant GBMF3849 and the Materials Research Science and Engineering Center (MRSEC) program of the National Science Foundation under Award Number DMR-1420073. A. Donev and B. Delmotte were supported in part by the National Science Foundation under award DMS-1418706. P. Chaikin was partially supported by NASA under Grant Number NNX13AR67G.
 
 \item[Contributions] M.D. performed the experiments. B.D. performed the simulations. M.Y. and S.S. synthesised the colloidal particles. M.D., B.D., A.D. and P.C. conceived the project, analysed the results and wrote the paper.
 
 \item[Competing Interests] The authors declare that they have no
competing financial interests.
 \item[Correspondence] Correspondence and requests for materials
should be addressed to Michelle Driscoll (mdriscoll@nyu.edu) or Blaise Delmotte (delmotte@courant.nyu.edu).
\end{addendum}

\newpage

\bibliography{Bib_nature}

\begin{thebibliography}{10}
\expandafter\ifx\csname url\endcsname\relax
  \def\url#1{\texttt{#1}}\fi
\expandafter\ifx\csname urlprefix\endcsname\relax\def\urlprefix{URL }\fi
\providecommand{\bibinfo}[2]{#2}
\providecommand{\eprint}[2][]{\url{#2}}

\bibitem{Anderson2002}
\bibinfo{author}{Anderson, V.~J.} \& \bibinfo{author}{Lekkerkerker, H.~N.}
\newblock \bibinfo{title}{Insights into phase transition kinetics from colloid
  science}.
\newblock \emph{\bibinfo{journal}{Nature}} \textbf{\bibinfo{volume}{416}},
  \bibinfo{pages}{811--815} (\bibinfo{year}{2002}).

\bibitem{Aranson2006}
\bibinfo{author}{Aranson, I.~S.} \& \bibinfo{author}{Tsimring, L.~S.}
\newblock \bibinfo{title}{Patterns and collective behavior in granular media:
  Theoretical concepts}.
\newblock \emph{\bibinfo{journal}{Reviews of modern physics}}
  \textbf{\bibinfo{volume}{78}}, \bibinfo{pages}{641} (\bibinfo{year}{2006}).

\bibitem{Marchetti2013}
\bibinfo{author}{Marchetti, M.} \emph{et~al.}
\newblock \bibinfo{title}{Hydrodynamics of soft active matter}.
\newblock \emph{\bibinfo{journal}{Reviews of Modern Physics}}
  \textbf{\bibinfo{volume}{85}}, \bibinfo{pages}{1143} (\bibinfo{year}{2013}).

\bibitem{Palacci2013}
\bibinfo{author}{Palacci, J.}, \bibinfo{author}{Sacanna, S.},
  \bibinfo{author}{Steinberg, A.~P.}, \bibinfo{author}{Pine, D.~J.} \&
  \bibinfo{author}{Chaikin, P.~M.}
\newblock \bibinfo{title}{Living crystals of light-activated colloidal
  surfers}.
\newblock \emph{\bibinfo{journal}{Science}} \textbf{\bibinfo{volume}{339}},
  \bibinfo{pages}{936--940} (\bibinfo{year}{2013}).

\bibitem{Bialke2015}
\bibinfo{author}{Bialk{\'e}, J.}, \bibinfo{author}{Speck, T.} \&
  \bibinfo{author}{L{\"o}wen, H.}
\newblock \bibinfo{title}{Active colloidal suspensions: Clustering and phase
  behavior}.
\newblock \emph{\bibinfo{journal}{Journal of Non-Crystalline Solids}}
  \textbf{\bibinfo{volume}{407}}, \bibinfo{pages}{367--375}
  (\bibinfo{year}{2015}).

\bibitem{Lu2013}
\bibinfo{author}{Lu, P.~J.} \& \bibinfo{author}{Weitz, D.~A.}
\newblock \bibinfo{title}{Colloidal particles: crystals, glasses, and gels}.
\newblock \emph{\bibinfo{journal}{Annu. Rev. Condens. Matter Phys.}}
  \textbf{\bibinfo{volume}{4}}, \bibinfo{pages}{217--233}
  (\bibinfo{year}{2013}).

\bibitem{Schwarz2012}
\bibinfo{author}{Schwarz-Linek, J.} \emph{et~al.}
\newblock \bibinfo{title}{Phase separation and rotor self-assembly in active
  particle suspensions}.
\newblock \emph{\bibinfo{journal}{Proceedings of the National Academy of
  Sciences}} \textbf{\bibinfo{volume}{109}}, \bibinfo{pages}{4052--4057}
  (\bibinfo{year}{2012}).

\bibitem{Vicsek2012}
\bibinfo{author}{Vicsek, T.} \& \bibinfo{author}{Zafeiris, A.}
\newblock \bibinfo{title}{Collective motion}.
\newblock \emph{\bibinfo{journal}{Physics Reports}}
  \textbf{\bibinfo{volume}{517}}, \bibinfo{pages}{71--140}
  (\bibinfo{year}{2012}).

\bibitem{Blake1974}
\bibinfo{author}{Blake, J.} \& \bibinfo{author}{Chwang, A.}
\newblock \bibinfo{title}{Fundamental singularities of viscous flow}.
\newblock \emph{\bibinfo{journal}{Journal of Engineering Mathematics}}
  \textbf{\bibinfo{volume}{8}}, \bibinfo{pages}{23--29} (\bibinfo{year}{1974}).

\bibitem{Lee1980}
\bibinfo{author}{Lee, S.} \& \bibinfo{author}{Leal, L.}
\newblock \bibinfo{title}{Motion of a sphere in the presence of a plane
  interface. {Part} 2. an exact solution in bipolar co-ordinates}.
\newblock \emph{\bibinfo{journal}{Journal of Fluid Mechanics}}
  \textbf{\bibinfo{volume}{98}}, \bibinfo{pages}{193--224}
  (\bibinfo{year}{1980}).

\bibitem{Huppert1982}
\bibinfo{author}{Huppert, H.~E.}
\newblock \bibinfo{title}{Flow and instability of a viscous current down a
  slope}.
\newblock \emph{\bibinfo{journal}{Nature}} \textbf{\bibinfo{volume}{300}},
  \bibinfo{pages}{427--429} (\bibinfo{year}{1982}).

\bibitem{Chandrasekhar1961}
\bibinfo{author}{Chandrasekhar, S.}
\newblock \bibinfo{title}{\emph{Hydrodynamic and Hydromagnetic Stability}}.
\newblock \emph{\bibinfo{journal}{\textup{International Series of Monographs on
  Physics, (Oxford, Clarendon, 1961)}}}  (\bibinfo{year}{1961}).

\bibitem{Sacanna2012}
\bibinfo{author}{Sacanna, S.}, \bibinfo{author}{Rossi, L.} \&
  \bibinfo{author}{Pine, D.~J.}
\newblock \bibinfo{title}{Magnetic click colloidal assembly}.
\newblock \emph{\bibinfo{journal}{Journal of the American Chemical Society}}
  \textbf{\bibinfo{volume}{134}}, \bibinfo{pages}{6112--6115}
  (\bibinfo{year}{2012}).

\bibitem{Bricard2013}
\bibinfo{author}{Bricard, A.}, \bibinfo{author}{Caussin, J.-B.},
  \bibinfo{author}{Desreumaux, N.}, \bibinfo{author}{Dauchot, O.} \&
  \bibinfo{author}{Bartolo, D.}
\newblock \bibinfo{title}{Emergence of macroscopic directed motion in
  populations of motile colloids}.
\newblock \emph{\bibinfo{journal}{Nature}} \textbf{\bibinfo{volume}{503}},
  \bibinfo{pages}{95--98} (\bibinfo{year}{2013}).

\bibitem{Sing2010}
\bibinfo{author}{Sing, C.~E.}, \bibinfo{author}{Schmid, L.},
  \bibinfo{author}{Schneider, M.~F.}, \bibinfo{author}{Franke, T.} \&
  \bibinfo{author}{Alexander-Katz, A.}
\newblock \bibinfo{title}{Controlled surface-induced flows from the motion of
  self-assembled colloidal walkers}.
\newblock \emph{\bibinfo{journal}{Proceedings of the National Academy of
  Sciences}} \textbf{\bibinfo{volume}{107}}, \bibinfo{pages}{535--540}
  (\bibinfo{year}{2010}).

\bibitem{Martinez2015}
\bibinfo{author}{Martinez-Pedrero, F.}, \bibinfo{author}{Ortiz-Ambriz, A.},
  \bibinfo{author}{Pagonabarraga, I.} \& \bibinfo{author}{Tierno, P.}
\newblock \bibinfo{title}{Colloidal microworms propelling via a cooperative
  hydrodynamic conveyor belt}.
\newblock \emph{\bibinfo{journal}{Physical review letters}}
  \textbf{\bibinfo{volume}{115}}, \bibinfo{pages}{138301}
  (\bibinfo{year}{2015}).

\bibitem{Grzybowski2000}
\bibinfo{author}{Grzybowski, B.~A.}, \bibinfo{author}{Stone, H.~A.} \&
  \bibinfo{author}{Whitesides, G.~M.}
\newblock \bibinfo{title}{Dynamic self-assembly of magnetized, millimetre-sized
  objects rotating at a liquid--air interface}.
\newblock \emph{\bibinfo{journal}{Nature}} \textbf{\bibinfo{volume}{405}},
  \bibinfo{pages}{1033--1036} (\bibinfo{year}{2000}).

\bibitem{Snezhko2016}
\bibinfo{author}{Snezhko, A.}
\newblock \bibinfo{title}{Complex collective dynamics of active torque-driven
  colloids at interfaces}.
\newblock \emph{\bibinfo{journal}{Current Opinion in Colloid \& Interface
  Science}} \textbf{\bibinfo{volume}{21}}, \bibinfo{pages}{65--75}
  (\bibinfo{year}{2016}).

\bibitem{Swan2007}
\bibinfo{author}{Swan, J.~W.} \& \bibinfo{author}{Brady, J.~F.}
\newblock \bibinfo{title}{Simulation of hydrodynamically interacting particles
  near a no-slip boundary}.
\newblock \emph{\bibinfo{journal}{Physics of Fluids (1994-present)}}
  \textbf{\bibinfo{volume}{19}}, \bibinfo{pages}{113306}
  (\bibinfo{year}{2007}).

\bibitem{Usabiaga2016}
\bibinfo{author}{Usabiaga, F.~B.} \emph{et~al.}
\newblock \bibinfo{title}{Hydrodynamics of suspensions of passive and active
  rigid particles: A rigid multiblob approach}.
\newblock \emph{\bibinfo{journal}{\textup{Preprint at
  http://arxiv.org/abs/1602.02170}}}  (\bibinfo{year}{2016}).

\bibitem{Burgers1974}
\bibinfo{author}{Burgers, J.}
\newblock \emph{\bibinfo{title}{The Nonlinear Diffusion Equation: Asymptotic
  Solutions and Statistical Problems}}.
\newblock Lecture series (\bibinfo{publisher}{Springer, Netherlands},
  \bibinfo{year}{1974}).

\bibitem{Beatus2009}
\bibinfo{author}{Beatus, T.}, \bibinfo{author}{Tlusty, T.} \&
  \bibinfo{author}{Bar-Ziv, R.}
\newblock \bibinfo{title}{Burgers shock waves and sound in a 2d microfluidic
  droplets ensemble}.
\newblock \emph{\bibinfo{journal}{Physical review letters}}
  \textbf{\bibinfo{volume}{103}}, \bibinfo{pages}{114502}
  (\bibinfo{year}{2009}).

\bibitem{Desreumaux2013}
\bibinfo{author}{Desreumaux, N.}, \bibinfo{author}{Caussin, J.-B.},
  \bibinfo{author}{Jeanneret, R.}, \bibinfo{author}{Lauga, E.} \&
  \bibinfo{author}{Bartolo, D.}
\newblock \bibinfo{title}{Hydrodynamic fluctuations in confined particle-laden
  fluids}.
\newblock \emph{\bibinfo{journal}{Physical review letters}}
  \textbf{\bibinfo{volume}{111}}, \bibinfo{pages}{118301}
  (\bibinfo{year}{2013}).

\bibitem{Goodman1999}
\bibinfo{author}{Goodman, J.} \& \bibinfo{author}{Miller, J.~R.}
\newblock \bibinfo{title}{Long-time behavior of scalar viscous shock fronts in
  two dimensions}.
\newblock \emph{\bibinfo{journal}{Journal of Dynamics and Differential
  Equations}} \textbf{\bibinfo{volume}{11}}, \bibinfo{pages}{255--277}
  (\bibinfo{year}{1999}).

\bibitem{Pouliquen1997}
\bibinfo{author}{Pouliquen, O.}, \bibinfo{author}{Delour, J.} \&
  \bibinfo{author}{Savage, S.}
\newblock \bibinfo{title}{Fingering in granular flows}.
\newblock \emph{\bibinfo{journal}{Nature}} \textbf{\bibinfo{volume}{386}},
  \bibinfo{pages}{816--817} (\bibinfo{year}{1997}).

\bibitem{Pan2001}
\bibinfo{author}{Pan, T.}, \bibinfo{author}{Joseph, D.} \&
  \bibinfo{author}{Glowinski, R.}
\newblock \bibinfo{title}{Modelling rayleigh--taylor instability of a
  sedimenting suspension of several thousand circular particles in a direct
  numerical simulation}.
\newblock \emph{\bibinfo{journal}{Journal of Fluid Mechanics}}
  \textbf{\bibinfo{volume}{434}}, \bibinfo{pages}{23--37}
  (\bibinfo{year}{2001}).

\bibitem{Lin2014}
\bibinfo{author}{Lin, T.}, \bibinfo{author}{Rubinstein, S.~M.},
  \bibinfo{author}{Korchev, A.} \& \bibinfo{author}{Weitz, D.~A.}
\newblock \bibinfo{title}{Pattern formation of charged particles in an electric
  field}.
\newblock \emph{\bibinfo{journal}{Langmuir}} \textbf{\bibinfo{volume}{30}},
  \bibinfo{pages}{12119--12123} (\bibinfo{year}{2014}).

\bibitem{Wysocki2009}
\bibinfo{author}{Wysocki, A.} \emph{et~al.}
\newblock \bibinfo{title}{Direct observation of hydrodynamic instabilities in a
  driven non-uniform colloidal dispersion}.
\newblock \emph{\bibinfo{journal}{Soft Matter}} \textbf{\bibinfo{volume}{5}},
  \bibinfo{pages}{1340--1344} (\bibinfo{year}{2009}).

\bibitem{Metzger2007}
\bibinfo{author}{Metzger, B.}, \bibinfo{author}{Nicolas, M.} \&
  \bibinfo{author}{Guazzelli, E.}
\newblock \bibinfo{title}{Falling clouds of particles in viscous fluids}.
\newblock \emph{\bibinfo{journal}{Journal of Fluid Mechanics}}
  \textbf{\bibinfo{volume}{580}}, \bibinfo{pages}{283--301}
  (\bibinfo{year}{2007}).

\bibitem{Lowen2010}
\bibinfo{author}{L{\"o}wen, H.}
\newblock \bibinfo{title}{Particle-resolved instabilities in colloidal
  dispersions}.
\newblock \emph{\bibinfo{journal}{Soft Matter}} \textbf{\bibinfo{volume}{6}},
  \bibinfo{pages}{3133--3142} (\bibinfo{year}{2010}).

\bibitem{Guazzelli2011}
\bibinfo{author}{Guazzelli, E.} \& \bibinfo{author}{Hinch, J.}
\newblock \bibinfo{title}{Fluctuations and instability in sedimentation}.
\newblock \emph{\bibinfo{journal}{Annual review of fluid mechanics}}
  \textbf{\bibinfo{volume}{43}}, \bibinfo{pages}{97--116}
  (\bibinfo{year}{2011}).

\bibitem{trackpy}
\bibinfo{author}{Allan, D.}, \bibinfo{author}{Caswell, T.},
  \bibinfo{author}{Keim, N.} \& \bibinfo{author}{van~der Wel, C.}
\newblock \bibinfo{title}{trackpy: Trackpy v0.3.2} (\bibinfo{year}{2016}).
\newblock \urlprefix\url{https://doi.org/10.5281/zenodo.60550}.

\end{thebibliography}


\begin{thebibliography}{1}
\expandafter\ifx\csname url\endcsname\relax
  \def\url#1{\texttt{#1}}\fi
\expandafter\ifx\csname urlprefix\endcsname\relax\def\urlprefix{URL }\fi
\providecommand{\bibinfo}[2]{#2}
\providecommand{\eprint}[2][]{\url{#2}}

\bibitem{Swan2007}
\bibinfo{author}{Swan, J.~W.} \& \bibinfo{author}{Brady, J.~F.}
\newblock \bibinfo{title}{Simulation of hydrodynamically interacting particles
  near a no-slip boundary}.
\newblock \emph{\bibinfo{journal}{Physics of Fluids (1994-present)}}
  \textbf{\bibinfo{volume}{19}}, \bibinfo{pages}{113306}
  (\bibinfo{year}{2007}).

\bibitem{Usabiaga2016}
\bibinfo{author}{Usabiaga, F.~B.} \emph{et~al.}
\newblock \bibinfo{title}{Hydrodynamics of suspensions of passive and active
  rigid particles: A rigid multiblob approach}.
\newblock \emph{\bibinfo{journal}{arXiv preprint arXiv:1602.02170}}
  (\bibinfo{year}{2016}).

\bibitem{Cebers2006}
\bibinfo{author}{C{\=e}bers, A.} \& \bibinfo{author}{Ozols, M.}
\newblock \bibinfo{title}{Dynamics of an active magnetic particle in a rotating
  magnetic field}.
\newblock \emph{\bibinfo{journal}{Physical Review E}}
  \textbf{\bibinfo{volume}{73}}, \bibinfo{pages}{021505}
  (\bibinfo{year}{2006}).

\bibitem{Sinn2011}
\bibinfo{author}{Sinn, I.} \emph{et~al.}
\newblock \bibinfo{title}{Magnetically uniform and tunable janus particles}.
\newblock \emph{\bibinfo{journal}{Applied Physics Letters}}
  \textbf{\bibinfo{volume}{98}}, \bibinfo{pages}{024101}
  (\bibinfo{year}{2011}).

\bibitem{Yan2015}
\bibinfo{author}{Yan, J.}, \bibinfo{author}{Bae, S.~C.} \&
  \bibinfo{author}{Granick, S.}
\newblock \bibinfo{title}{Rotating crystals of magnetic janus colloids}.
\newblock \emph{\bibinfo{journal}{Soft Matter}} \textbf{\bibinfo{volume}{11}},
  \bibinfo{pages}{147--153} (\bibinfo{year}{2015}).

\bibitem{Melle2003}
\bibinfo{author}{Melle, S.} \& \bibinfo{author}{Martin, J.~E.}
\newblock \bibinfo{title}{Chain model of a magnetorheological suspension in a
  rotating field}.
\newblock \emph{\bibinfo{journal}{The Journal of chemical physics}}
  \textbf{\bibinfo{volume}{118}}, \bibinfo{pages}{9875--9881}
  (\bibinfo{year}{2003}).

\bibitem{Du2016}
\bibinfo{author}{Du, D.}, \bibinfo{author}{Hilou, E.} \&
  \bibinfo{author}{Biswal, S.~L.}
\newblock \bibinfo{title}{Modified mason number for charged paramagnetic
  colloidal suspensions}.
\newblock \emph{\bibinfo{journal}{Physical Review E}}
  \textbf{\bibinfo{volume}{93}}, \bibinfo{pages}{062603}
  (\bibinfo{year}{2016}).

\end{thebibliography}

\newpage

\begin{figure}[t]
\includegraphics[width=0.9\textwidth]{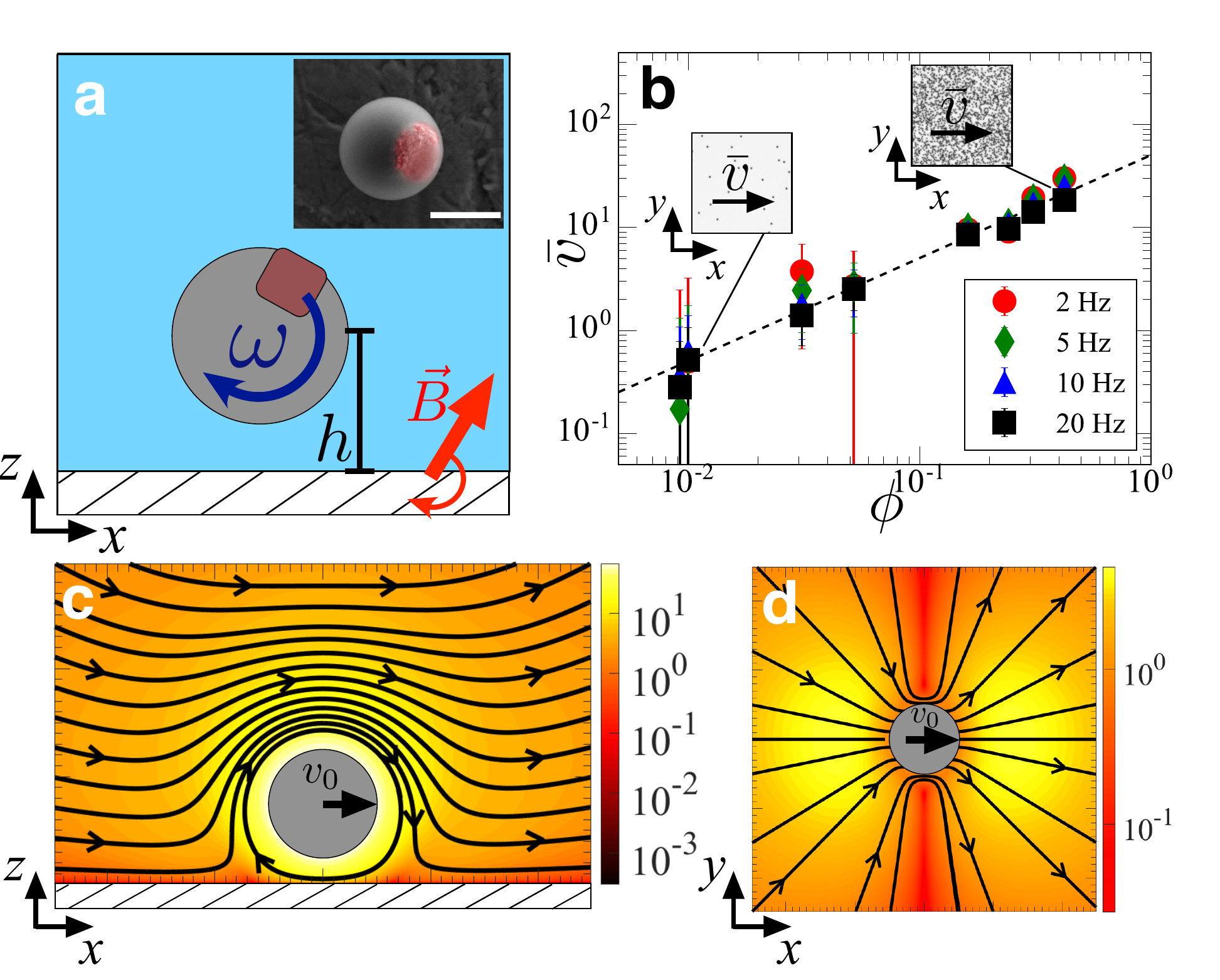} 
\caption{\textbf{Strong velocity enhancement due to collective effects.}  \textbf{a}, SEM and schematic of the polymer colloids with an embedded magnetic cube, indicated in red (scale bar is 1 $\mu$m).  A rotating magnetic field, $\vec{B}$, with angular frequency $\omega$ directs particle motion.
\textbf{b}, Normalised particle velocity \MD{measured in experiments,} $\bar{v} = (v-v_0)/v_0$, vs. area fraction, $\phi$, at fixed field strength, $B_0$ = 3 mT, for various frequencies $f = \omega/2\pi$. Insets show pictures of the system at the highest and lowest $\phi$.  The best linear fit (dashed line) shows a strong dependence of $\bar{v}$ on $\phi$, \MD{ $\bar{v} = \alpha\phi$, where $\alpha = 50\pm 2$.}  \textbf{c, d}, Calculated streamlines around a rotating particle ($f$ = 10 Hz, $h$ = 0.98 $\mu$m) in the plane perpendicular (\textbf{c}) and parallel (\textbf{d}) to the rotation direction.  Flow velocity is normalised by single particle translation velocity $v_0$ and its magnitude is shown by the colour bar.}
\end{figure}

\newpage

\begin{figure}[t]
\includegraphics[width=0.9\textwidth]{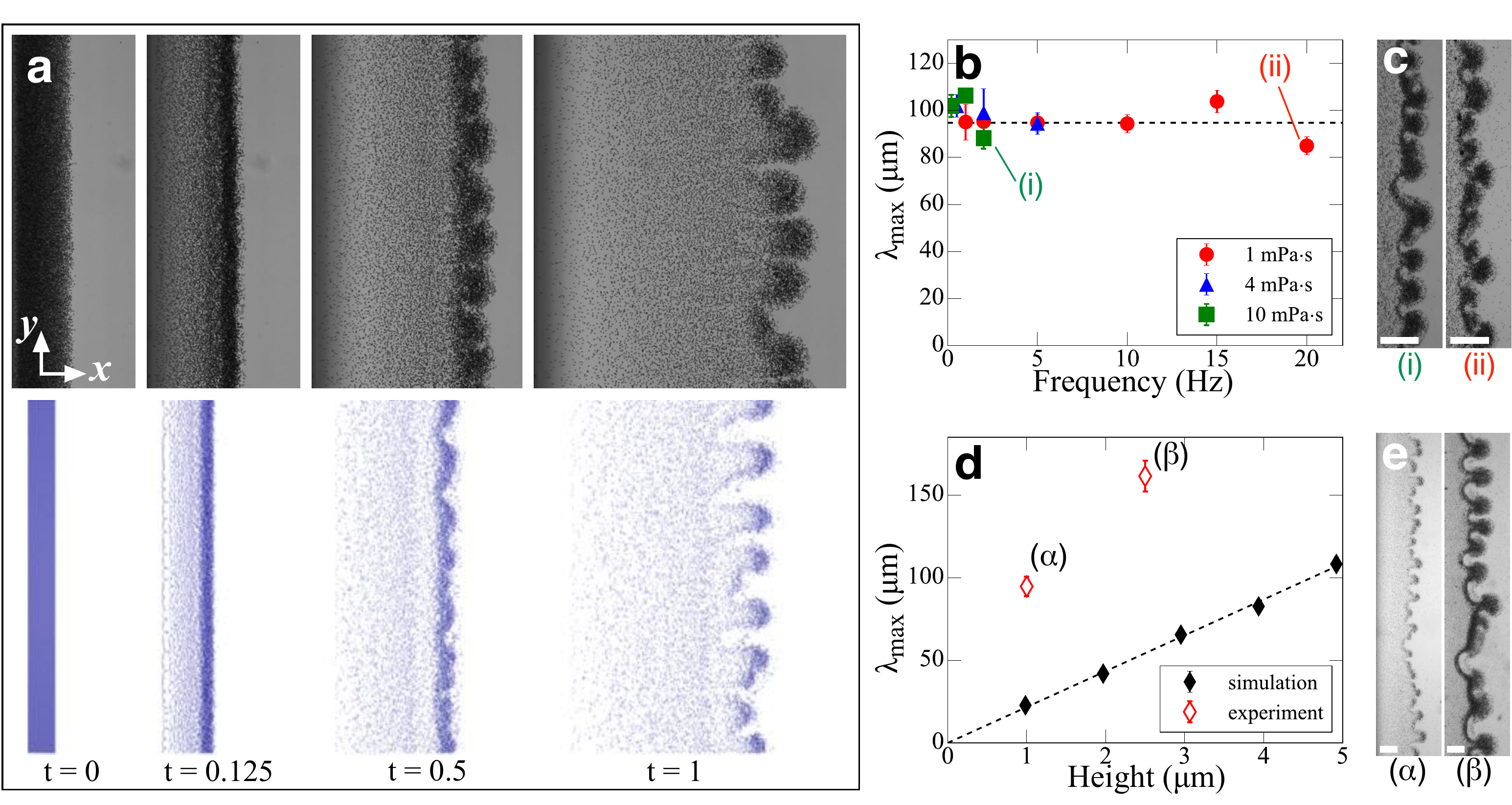} 
\caption{\textbf{Fingering instability.} \textbf{a}, Instability in the experiment (above) and simulation (below) at the same relative points in time (in the $xy$-plane). Labels indicate scaled time (arbitrary units), and all images are 0.55 mm tall. \textbf{b}, Experimental data of instability wavelength, $\lambda_{\max}$ vs. $f$, for three different fluid viscosities (different symbols), dashed line indicates mean $\lambda_{\max}$.  \textbf{c}, Experimental images corresponding to (i) $\eta$ = 1 mPa$\cdot$s and (ii) $\eta$ = 10 mPa$\cdot$s, scale bar indicates 100 $\mu$m. \textbf{d}, Experimental and simulation data of $\lambda_{\max}$ vs. $h$, dashed line indicates best linear fit \fixed{to the simulation data}.   \textbf{e}, Experimental images \fixed{for gravitational heights} ($\alpha$) $h$ = 1.0 $\mu$m and ($\beta$) $h$ = 2.5 $\mu$m, scale bar indicates 100 $\mu$m.}
\label{instability}
\end{figure}

\newpage

\begin{figure}[t]
\includegraphics[width=0.9\textwidth]{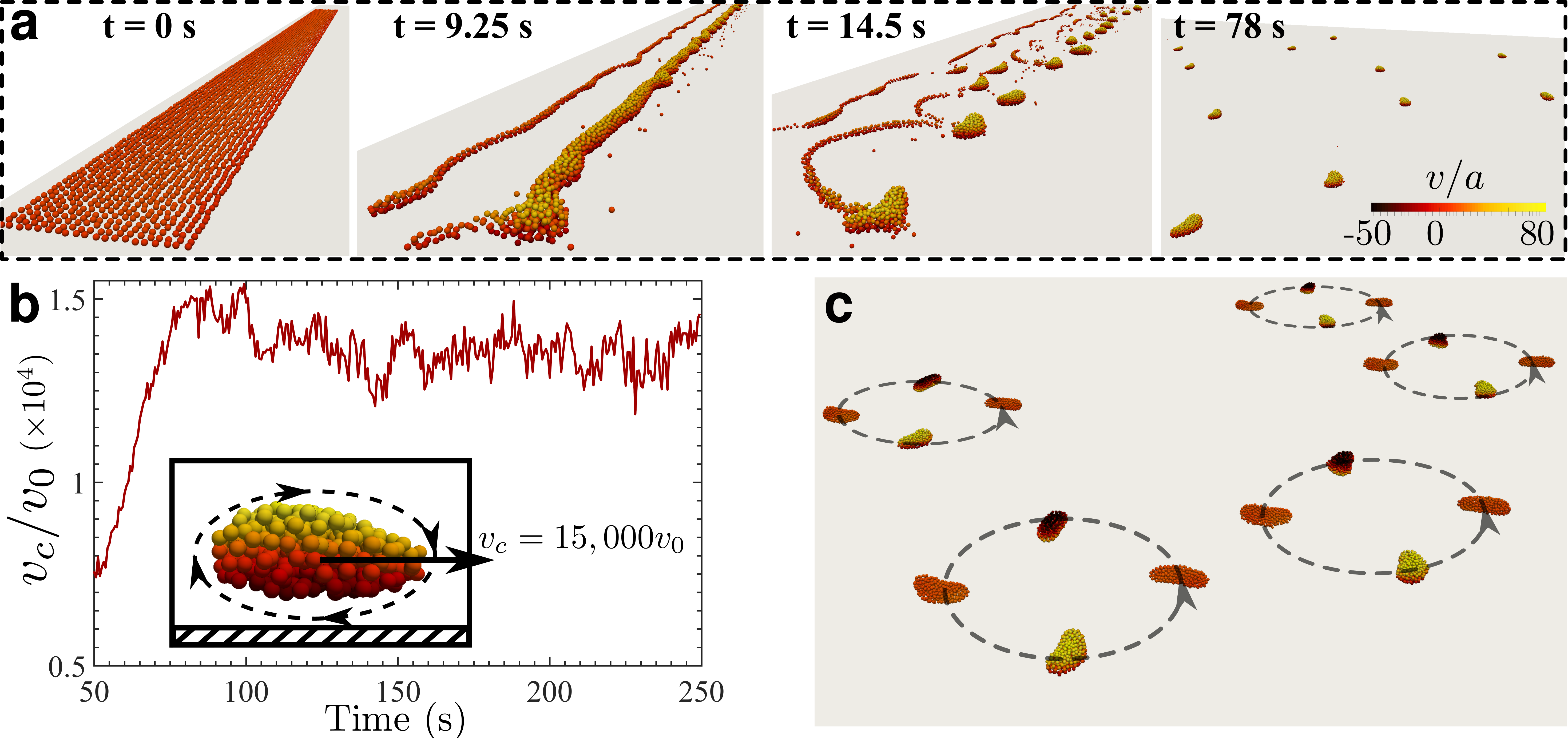} 
\caption{\textbf{Self-sustained critters.} \textbf{a}, Time evolution of the instability. Color bar indicates translational velocity in radii/s and the camera position changes dynamically. $t$  = 0 s: 8000 rollers are initially randomly distributed on a strip.  $t$  = 9.25 s: a compact front appears and starts to destabilise. $t$  = 14.5 s: the fingertips start to detach from the front and form critters. $t \geq$ 78 s: the critters reach a stable steady state in which they translate autonomously at a constant speed. \textbf{b}, Time evolution of normalised translational velocity of a critter, $v_c/v_0$, where $v_0$ is the velocity of a roller at the same centre of mass height. Inset: side view of the critter at $t$  = 78 s, as indicated by the arrows particles rotate about the centre of mass of the critter. \textbf{c}, Circular periodic trajectory of five critters rotating counter-clockwise in response to slowly changing the axis of rotation. The period of the trajectory is $T$  = 40 s and a frame is shown every $T/4$.} 
\label{critters}
\end{figure}

\newpage

\begin{figure}[t]
\includegraphics[width=0.9\textwidth]{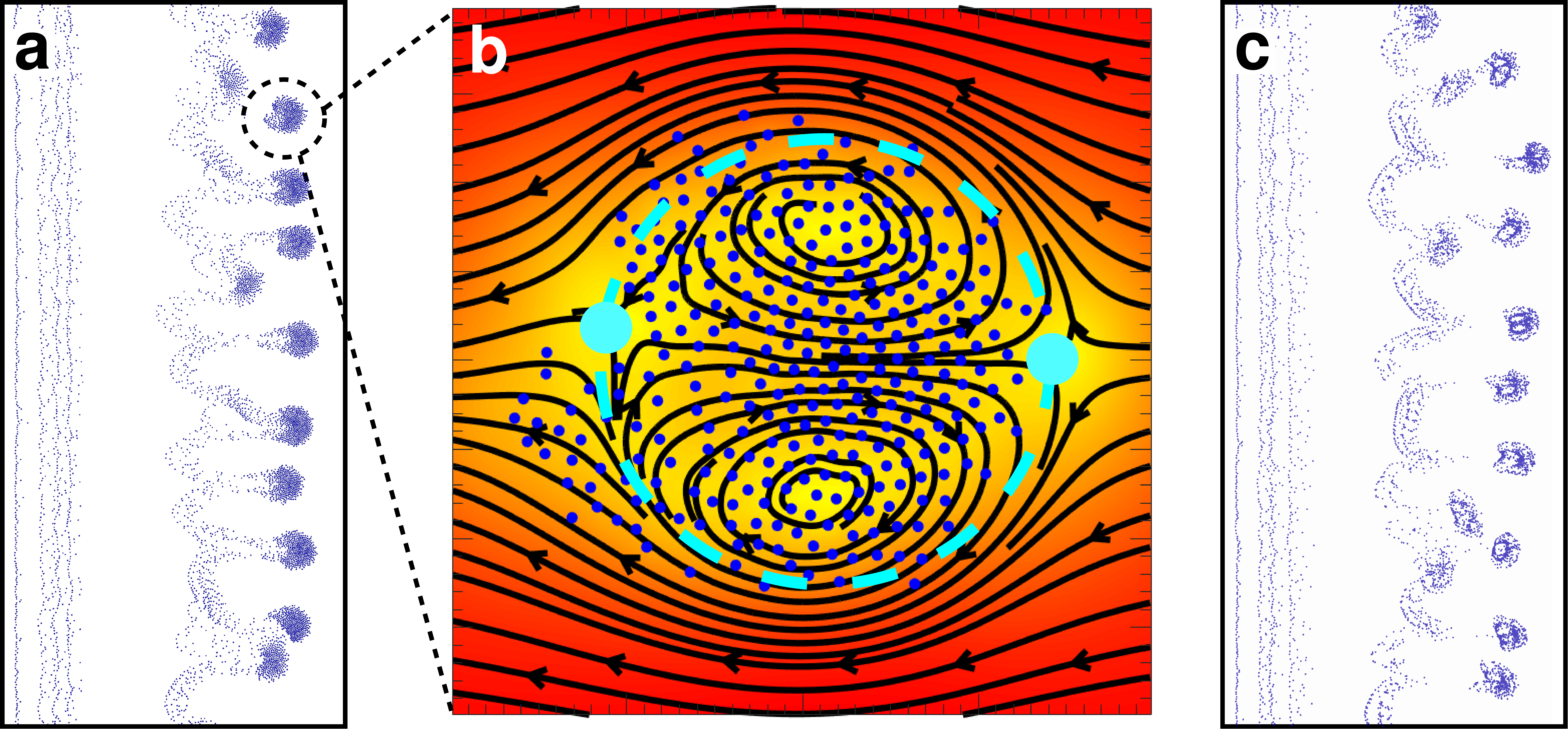} 
\caption{\textbf{Instability \& clustering controlled by hydrodynamics.} Simulations with rollers restricted to a plane $z=3.94$ $\mu$m, at time $t=74$ s. \textbf{a}, Simulations include both steric interactions and finite size effects. \textbf{b}, Flow field in the frame moving with a cluster. Blue dots indicate roller positions, the dashed cyan line circles the  recirculation zone where motion is self-sustained, and the cyan dots show the  stagnation points. \textbf{c}, Purely hydrodynamic simulation of rotlet singularities with no steric repulsion reproduce the instability.} 
\label{2D_pure_hydro}
\end{figure}

\newpage

\end{document}


\title{\huge{Supplementary Methods}}
\maketitle

\section{Rotation Mechanism}

\begin{figure}[h!]
\includegraphics[width=0.85\textwidth]{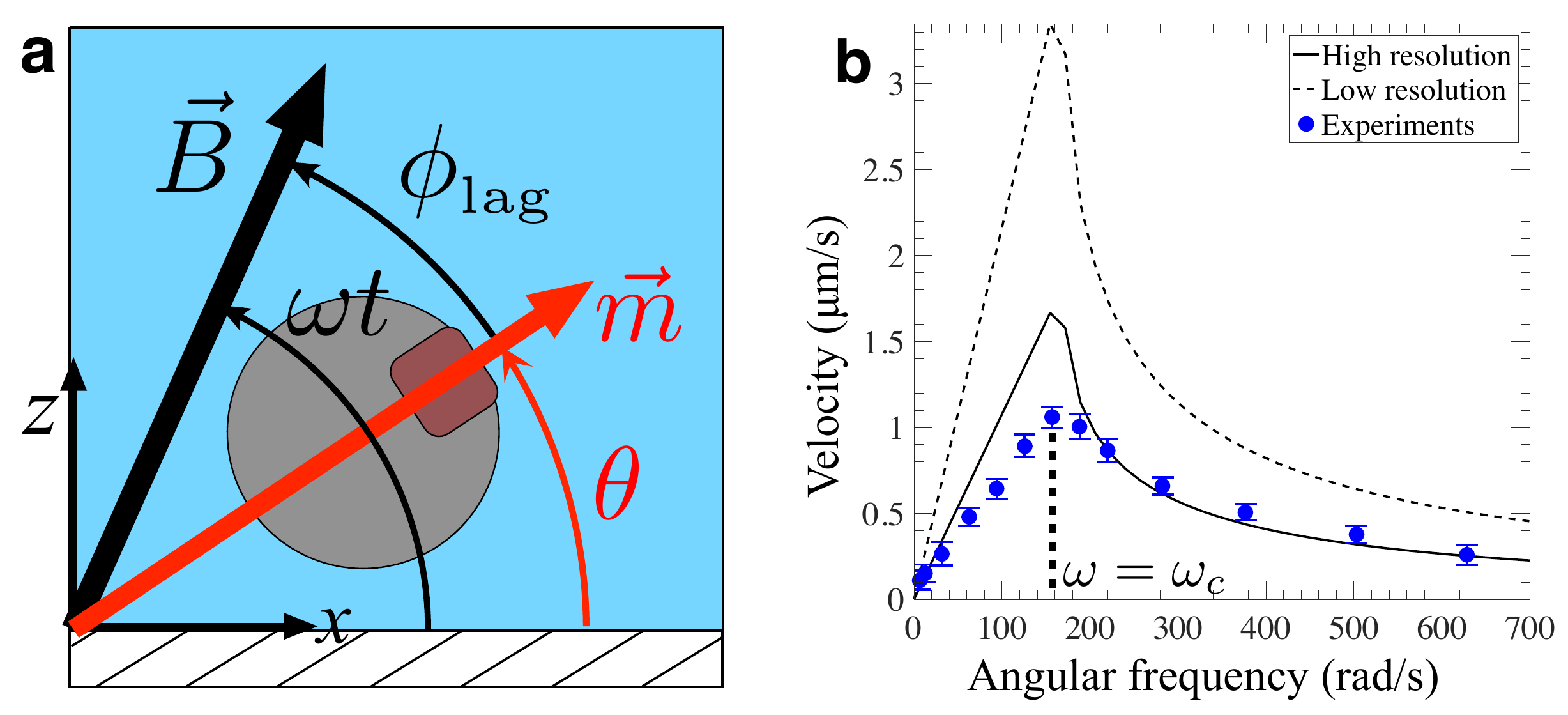} 
\begin{centering}
\caption{\MB{(a) Coordinate system of the rotating particle. In steady state, the particle rotates synchronously with the applied magnetic field, i.e.\ the phase lag $\phi_{\mbox{\tiny{lag}}}$ is constant.    (b)  Particle velocity vs.\ angular frequency.  In the experiments, the magnetic field strength is 2.94 mT and the area fraction is $\phi = 10^{-4}$. In the simulations, a single particle is considered. The lines show simulations for different levels of accuracy in modeling hydrodynamic interactions with the wall.  The simulations labeled ``High resolution" use a well-resolved ``multi-blob" approach, and the simulations labeled ``Low resolution" use the minimally-resolved model, see main text and Methods for details.  In all cases, above  $\omega_c$, the particle no longer rotates synchronously with the field, leading to a decrease in velocity.
}}
\end{centering}
\end{figure}

\MB{The particles used in this study have a permanent magnetic moment, ($\lvert \vec{m}\rvert \sim 5\cdot 10^{-16}$ Am$^2$).  When a magnetic field rotating at frequency $\omega$ is applied to the sample, the particles will rotate synchronously with the field if the field strength is high enough and the frequency is low enough so that the applied magnetic torque can overcome the viscous torque exerted on the particle by the fluid. This rotation, when near a wall, leads to particle translation, as discussed in the main text.

To understand the rotation rate of the particles in the applied field, one must compare the magnetic torque exerted by the applied field to the viscous torque from the fluid.  The magnetic torque on the particle is given by $\tau_{m} = \|\vec{m} \times \vec{B}\| = mB\sin{\phi_{\mbox{\tiny{lag}}}}$, where $B$ is the applied field and $\phi_{\mbox{\tiny{lag}}} = \omega t - \theta$ is the phase lag between the magnetic moment and the external field (see SI Figure 1a).  The viscous torque is given by $\tau_{\nu} = 8\pi\eta a^3f(a/h)\dot{\theta}$, where $\eta$ is the fluid viscosity, $a$ is the particle radius, $h$ is the height of the particle above the surface, and $f(a/h)$ is a function that describes the hydrodynamic interaction with the wall \Donev{($f(a/h)= \left(1-\tfrac{15}{48}\tfrac{a^3}{h^3}\right)^{-1}$ to leading order \cite{Swan2007}) that we numerically estimate using a well-resolved model of 2562 blobs \cite{Usabiaga2016}}.  Imposing a torque-free condition (as appropriate for Stokes flow), one finds that \cite{Cebers2006,Sinn2011,Yan2015}
\begin{equation}
    \dot{\phi}_{\mbox{\tiny{lag}}}(t) = \omega - \omega_c\sin{\phi_{\mbox{\tiny{lag}}}(t)}
\end{equation}
where the critical frequency is given by 
\begin{equation}
\omega_c = mB/(8\pi\eta a^3f(a/h)). 
    \label{omegaC}
\end{equation}
If $\omega/\omega_c\le1$, the system has a steady state solution given by $\phi_{\mbox{\tiny{lag}}}^\infty = \sin^{-1}(\omega/\omega_c)$, and the roller rotates synchronously with the magnetic field with a fixed phase lag. 
If $\omega > \omega_c$, no steady solution to Eq. 1 exists, i.e.\ the lag angle is now a periodic function of time where the particle rotation ``slips" relative to the rotating field once per period \cite{Cebers2006}.

The response of the magnetic particles used in this study is illustrated in SI Figure 1b.  The experiments at fixed area fraction ($\phi = 10^{-4}$) and field strength ($B = 2.94$ mT), show that the particle  velocity increases with rotation frequency until it reaches a peak value at $\omega_c = 170$ rad/s.  The decrease in  velocity for $\omega > \omega_c$ indicates that the particles are no longer able to follow the field.     Using Eq.\ \eqref{omegaC}, the value of $\lvert \vec{m}\rvert$ calculated from the measured value of $\omega_c$ is $4.5 \times 10^{-16}$ Am$^2$, which is in agreement with the value estimated from bulk hematite properties, $6.0$ $\times 10^{-16}$ Am$^2$. 

The lines in SI Figure 1b show \modified{simulation estimates for the particle velocity averaged over a period of rotation of the magnetic field and averaged over the Gibbs-Boltzmann distribution of heights $h$,} with the gravitational height measured \MB{  using the diffusion coefficient from}  experiments ($h \approx 1.0$  $\mu$m)\Donev{, and the magnetic moment set to $\lvert \vec{m}\rvert=5 \times 10^{-16}$ A$\cdot$m$^2$}. The low resolution calculation corresponds to the minimally-resolved one used in the large scale simulations \cite{Swan2007} \Donev{and represents the sphere using a single regularized rotlet and stokeslet (blob)}. The high resolution calculation corresponds to the well-resolved ``multi-blob" model used to obtain the flow fields around in a single roller in Figure 1 in the main text (see Methods for details), and represents the surface of the sphere as a rigid cluster of 2562 blobs. As shown in the Figure, the low resolution calculation clearly overestimates the velocity but captures the qualitative trends of the experimental curve. As the resolution increases, the velocity approaches the experimental result. A perfect match is not expected between the resolved simulations and experiments because of the sensitivity of the speed to the exact values of $a$ and $h$ ($v_0 \sim (a/h)^4$ to leading order), which are only known to within $10\%$.
}

\section{Mason Number}

The Mason number characterises the relative strength of viscous forces to magnetic forces.  We define the Mason number as \cite{Melle2003,Du2016}
\begin{equation*}
    \mbox{Ma} = \dfrac{144\pi\eta f}{\mu_0 M_p^2},
\end{equation*}
where $\eta$ is the dynamic viscosity of the suspending fluid, $f$ is the frequency of the rotating magnetic field, $\mu_0 = 4\pi \times 10^{-7}$ N/A$^2$ is the vacuum permeability, and $M_p$ is the magnetisation per particle.  We change the Mason number in this system by using colloids with much stronger magnetic interactions,  super-paramagnetic dynabeads (Fisher Scientific), see main text and SI Movie 9.  For the dynabeads, $M_p$ = $\chi H_0$ = 2292 A/m ($\chi$ = volumetric susceptibility, $H_0$ = applied field), giving Ma = 0.69 for a field strength of 3 mT.  For the TPM/hematite particles, which are permanently magnetised, $M_p$ = 85 A/m, giving Ma = 500.

\section{Characterisation of the fingering instability}
To extract the most unstable wavelength, the same Fourier analysis is performed on the density profiles in the simulations and experiments. In brief, the images are windowed to chose the region around the instability, and then the particle relative density in this region is extracted by taking the mean intensity of this windowed image averaged in the direction perpendicular to the front.  The Fourier spectrum of the resulting one-dimensional average density is then computed. 

In a linear instability at early times, the amplitude of each mode grows exponentially in time. To identify the fastest growing mode in our instability, the growth rate of each wavelength is first extracted as follows.  For each wavelength, the amplitude at early times (when the wave amplitude is smaller than the wavelength) is fit to an exponential function, $A_{\sigma}(t) = A_0\exp(\sigma t)$.  The value of $\sigma$ for all wavelengths is then examined to identify the fastest growing (largest $\sigma$) wavelength, $\lambda_{\max}$.  The results are averaged over 5 -- 10 experimental runs or 10 numerical realizations (see SI Figure 2).  The value of $\lambda_{\max}$ is determined by fitting a parabola to the peak in the averaged data over a narrow range centred at the maximum.

\begin{figure}[h!]
\includegraphics[width=0.85\textwidth]{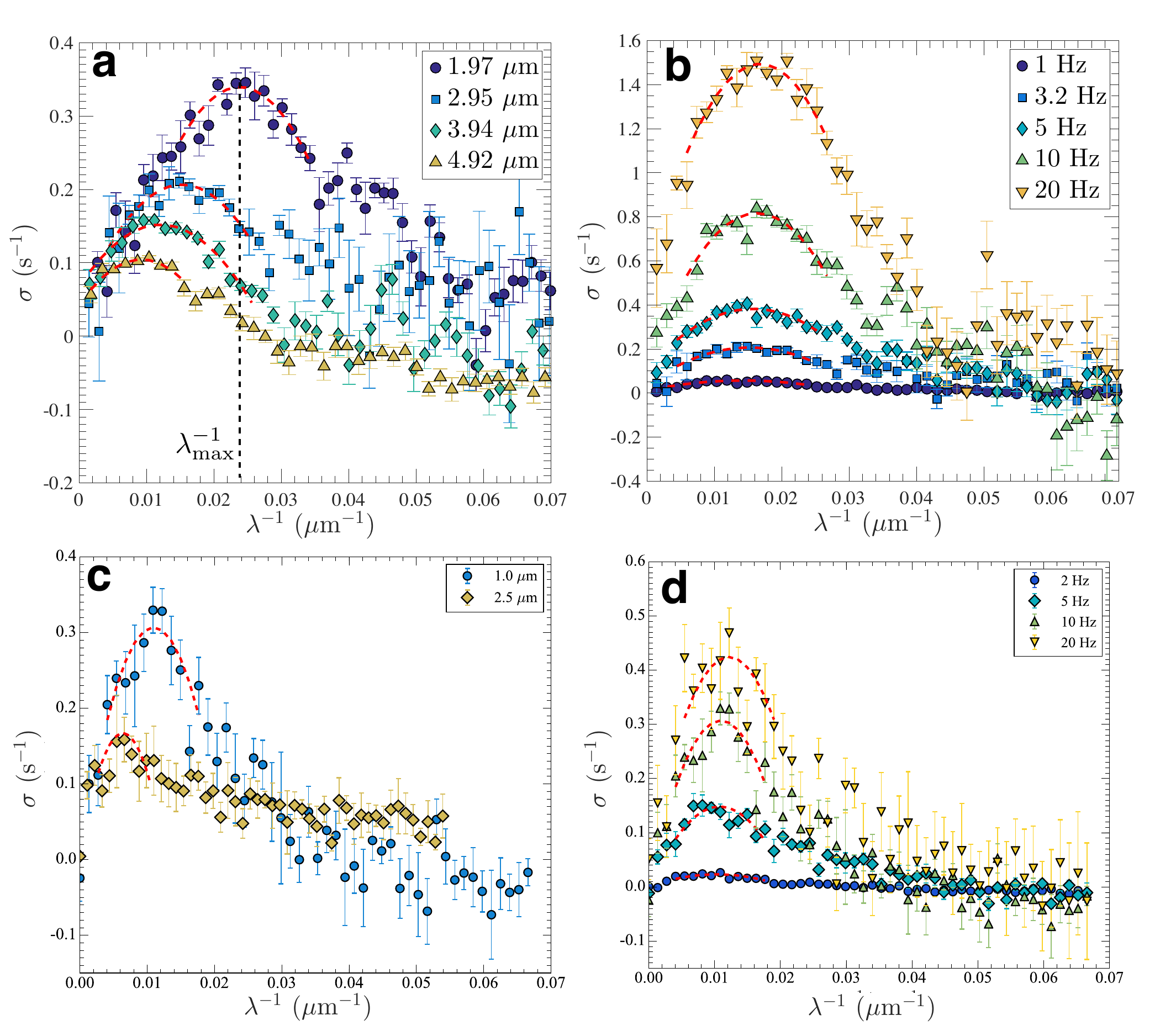} 
\begin{centering}
\caption{Growth rate $\sigma$ vs.\ $\lambda^{-1}$ in both simulations (a,b) and experiments (c,d). Red dashed lines indicate parabolic fits used to identify $\lambda_{\max}$. (a) Simulations with fixed frequency, $f = 3.2$ Hz, growth rate $\sigma$ vs.\ $\lambda^{-1}$ for several heights: $h = 1.97$ -- $4.92$ $\mu$m. $\lambda_{\max}$ is shown for $h = 1.97$ $\mu$m. (b) Simulations with fixed particle height, $h = 3.94$ $\mu$m, growth rate $\sigma$ vs.\ $\lambda^{-1}$ for various rotation frequencies, $f = $ 1 -- 20 Hz.  (c) Experiments at fixed frequency, $f =$ 10 Hz, growth rate $\sigma$ vs.\ $\lambda^{-1}$ for two equilibrium heights, $h = 1.0 $ $\mu$m and $h =$ 2.5 $\mu$m. (d) Experiments at fixed equilibrium height, $h =$ 1.0 $\mu$m, $\sigma$ vs.\ $\lambda^{-1}$ for several various rotation frequencies, $f =$ 2 -- 20 Hz.}
\end{centering}

\end{figure}

\newpage

\section{Transport of passive particles}

\MB{The dynamics of this system is dominated by advective flows, see Figure 1 in the main text.  Thus, if passive particles, i.e.\ ones which \emph{do not} rotate in the applied field are added to a suspension of rollers, they can be advected in the flow.  SI Figure 3 demonstrates three possible routes for particle transport: transport via the fingering instability, transport via critters, and transport via a homogeneous suspension of rollers.  In all three cases, the strong flows created by the rollers can be used to transport passive (non-magnetic) particles.  In the experiments (SI Figure 3a,c), large ($a$ = 14 $\mu$m) polystyrene particles are used for passive transport.  In the simulations (SI Figure 3b), passive particles are simply ones which are not prescribed to rotate.  

Rollers can also be used for guided transport, as illustrated in SI Figure 3c.  Here, the rolling particles were guided by changing the orientation of the rotating magnetic field during the experiment.  The ability to transport passive particles over macroscopic distances, as well as to guide them, suggests that rollers are a promising candidate system for controlling both microscopic mixing and large scale transport.}

\begin{figure}[h!]
\includegraphics[width=0.7\textwidth]{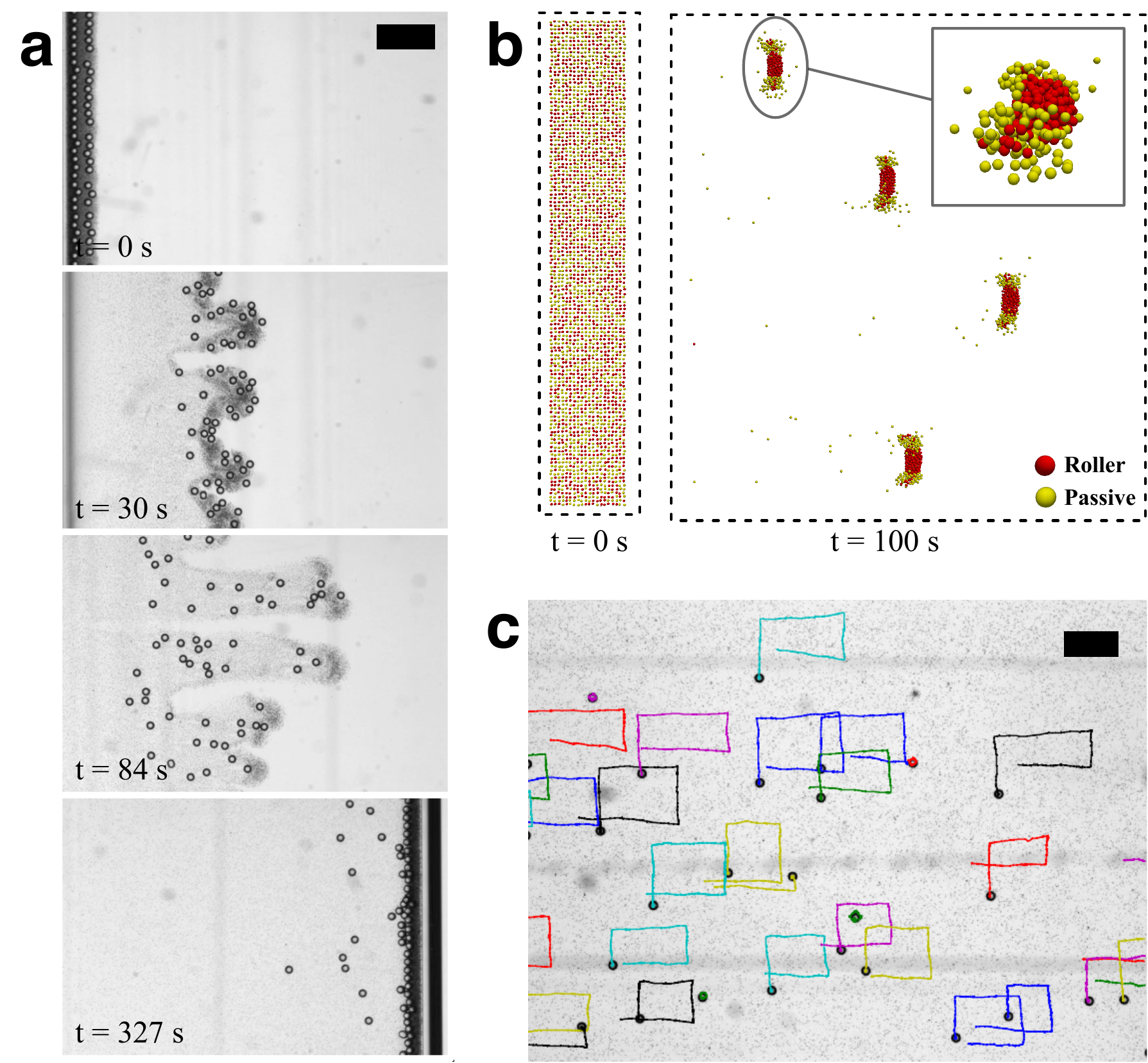} 
\begin{centering}
\caption{\MB{(a) Transport via fingering instability.  Experiment: the fingers resulting from the instability advect passive particles across a 2 mm capillary tube.  The passive particles are large ($a = 14 \mu$m), and the scale bar indicates 200 $\mu$m. (b) Transport via critters.  Simulation: starting with a 50:50 mixture of passive and active particles  results in the formation of critters transporting passive particles\modified{, which cluster on the lateral sides of the critters}. (c) Experiment: transport via a homogeneous suspension.  Rollers can be guided by changing the direction of the applied rotating field during the experiment.  Colored lines indicate the tracks of the passive particles (polystyrene, $a = 14 \mu$m) followed over the course of the experiment.  The total experimental time was 460 seconds, and the scale bar indicates 200 $\mu$m.}}
\end{centering}
\end{figure}

\newpage

\bibliographystyle{naturemag}
\bibliography{Bib_SI}